\documentclass{PoS}

\usepackage{amsmath}
\usepackage{comment}

\def\beq{\begin{equation}}
\def\eeq{\end{equation}}
\def\bea{\begin{eqnarray}}
\def\eea{\end{eqnarray}}
\def\bec{\begin{center}}
\def\eec{\end{center}}
\def\nn{\nonumber}

\newcommand{\cN}{{\cal N}}
\newcommand{\Tr}{{\rm Tr\;}}
\newcommand{\lambdab}{\overline{\lambda}}

\newcommand{\cA}{{\cal A}}
\newcommand{\cAb}{{\overline{\cal A}}}
\newcommand{\cF}{{\cal F}}
\newcommand{\cFb}{{\overline{\cal F}}}
\newcommand{\cD}{{\cal D}}
\newcommand{\cDb}{{\overline{\cal D}}}
\newcommand{\cP}{{\cal P}}
\newcommand{\cQ}{{\cal Q}}
\newcommand{\cU}{{\cal U}}
\newcommand{\cUb}{{\overline{\cal U}}} 

\newcommand{\bn}{{\bf n}}

\newcommand{\hatbmu}{\widehat{\boldsymbol {\mu}}}

\newcommand{\hf}{\frac{1}{2}}
\newcommand{\qtr}{\frac{1}{4}}

\title{\vspace{-2.5cm}{\small \normalfont \hfill DAMTP-2016-69}\vspace{3cm}\\ 5D Maximally Supersymmetric Yang-Mills on the Lattice}

\ShortTitle{5D Maximally Supersymmetric Yang-Mills on the Lattice}

\author{\speaker{Anosh Joseph} \\
Department of Applied Mathematics and Theoretical Physics (DAMTP) \\
Centre for Mathematical Sciences \\
University of Cambridge \\
Cambridge, CB3 0WA, UK \\
{} \\
International Centre for Theoretical Sciences (ICTS) \\
Tata Institute of Fundamental Research \\ 
Shivakote, Hesaraghatta \\
Bangalore 560089, India \\
{} \\
E-mail: \email{anosh.joseph@icts.res.in}}

\abstract{We provide details of the lattice construction of five-dimensional maximally supersymmetric Yang-Mills theory. The lattice theory is supersymmetric, gauge invariant and free from spectrum doublers. Such a supersymmetric lattice formulation is interesting as it can be used for non-perturbative explorations of the five-dimensional theory, which has a known gravitational dual.}

\FullConference{The 34th International Symposium on Lattice Field Theory,\\24-30 July, 2016\\University of Southampton, Southampton, UK}

\begin{document}

\section{Introduction}
\label{sec:intro}

Five-dimensional maximally supersymmetric Yang-Mills (MSYM) theory has generated a lot of interest in the recent years \cite{Douglas:2010iu, Tachikawa:2011ch, Kallen:2012cs, Kallen:2012va, Kim:2012ava, Kallen:2012zn, Bern:2012di, Minahan:2013jwa}. This theory takes part in the AdS-CFT correspondence and its finite temperature properties have been investigated recently in the planar limit. Both the gauge theory and supergravity calculations show that the free energy has $N^3$ scaling behavior in the large-$N$ limit \cite{Klebanov:1996un}. The five-dimensional gauge theory has a UV completion in the six-dimensional $(2, 0)$ superconformal theory. It was conjectured recently \cite{Douglas:2010iu} that 5D MSYM captures all the degrees of freedom of the parent six-dimensional theory suggesting that these two theories are the same.

We can learn a lot about the six-dimensional $(2,0)$ theory through the AdS-CFT correspondence, where the $(2, 0)$ theory is conjectured to be dual to M-theory (or supergravity) on $AdS_7 \times S^4$ background \cite{Maldacena:1997re}. Supergravity calculations reveal that the free energy of the $(2, 0)$ theory has $N^3$ scaling behavior \cite{Klebanov:1996un}. One can compactify the $(2, 0)$ theory on a circle to obtain the 5D MSYM. Since the $N^3$ behavior remains in the supergravity dual after compactification, one might expect to find some indication of it in the 5D MSYM. In Ref. \cite{Kallen:2012zn} the $N^3$ scaling behavior has been reproduced in the five-dimensional theory, at large 't Hooft coupling, through calculations based on large-$N$ matrix model, agreeing with the supergravity prediction.

In the five-dimensional theory the square of the gauge coupling has the dimension of length and a simple power counting suggests that the theory is not perturbatively renormalizable. The 6D theory is believed to be UV finite suggesting that the 5D theory is also UV finite in order for the conjecture in Ref. \cite{Douglas:2010iu} to hold. In Ref. \cite{Bern:2012di} the authors computed the six-loop four-point correlation function of the 5D theory in the planar limit and showed that it is divergent. This indicates that the connection between the five- and six-dimensional theories is not so straightforward and the claim that these two theories are equivalent may not be true. Though this conjecture may not be true due to the existence of the UV divergences in 5D MSYM, one can still expect that the supersymmetric observables of these two theories match. For example, one can compute the expectation values of the Wilson loop in the 5D MSYM, and through the relation between the compactification radius and Yang-Mills coupling, one could compute the Wilson surface in the $(2, 0)$ theory.

We provide details of the construction of a supersymmetric lattice for five-dimensional maximally supersymmetric Yang-Mills theory. See Ref. \cite{Lattice_Susy} for recent developments in the area of lattice supersymmetry. 

\section{Continuum 5D MSYM}	
\label{sec:5d_n2_msym}

The five-dimensional MSYM theory can be obtained by dimensional reduction of $\cN=1$ SYM in ten dimensions. The five-dimensional theory resulting from dimensional reduction has $\cN=2$ supersymmetry, with a Euclidean rotation group $SO_E(5)$ and an R-symmetry group $SO_R(5)$. The action of the five-dimensional theory is
\bea
S_5 &=& \frac{1}{g_5^2} \int d^5x ~\Tr \Big(\qtr F_{mn} F_{mn} + \hf D_m \phi_j D_m \phi_j + \qtr [\phi_j, \phi_k] [\phi_j, \phi_k] \nn \\
&& \quad \quad \quad \quad \quad \quad \quad \quad \quad -i\lambdab^{aX}(\gamma^m)_a^{~~b} D_m \lambda_{bX} - \lambdab^{aX} (\gamma^j)_X^{~Y} [\phi_j, \lambda_{aY}] \Big),
\eea  
where the spinor indices $(a,b)$ correspond to the Euclidean rotation group and the indices $(X,Y)$ that of the R-symmetry group.

Since the R-symmetry group of the five-dimensional theory is as large as its Euclidean spacetime rotation group we can {\it maximally twist} this theory to obtain a twisted version of the theory. In five dimensions the twisting process is unique and it leads to a twisted theory with the B-model type twist. The topologically twisted version of this theory has been constructed in Ref. \cite{Geyer:2002gy} and we will follow the same procedure for the twisted theory in the continuum. In order to obtain the twisted version of this theory we take the new rotation group of the theory to be the diagonal subgroup of $SO_E(5) \times SO_R(5)$. This amounts to identifying the spinor indices, $a$ and $X$, and the spacetime and internal symmetry indices $m$ and $j$. 

The 5D theory has the following action after twist
\bea
S_5 &=& \frac{1}{g_5^2} \int d^5x ~\Tr \Big(\qtr \cFb_{mn} \cF_{mn} - d D_m \phi_m - \hf d^2 - i \chi_{mn} \cD_m \psi_n - i \psi_m \cDb_m \eta \nn \\
&& \quad \quad \quad \quad \quad \quad \quad \quad \quad \quad - \frac{i}{8} \epsilon_{mncde} \chi_{de} \cDb_c \chi_{mn} \Big),
\eea  
where $\phi_m$ is now promoted to a five-dimensional vector field after twisting and $\eta$, $\psi_m$ and $\chi_{mn}$ are twisted fermions. It is natural to combine the two vector fields $A_m$ and $\phi_m$ to form a complexified gauge field: $\cA_m = A_m + i\phi_m$ and $\cAb_m = A_m - i\phi_m$. The complexified field strengths appearing in the above action are defined as
\bea
\cF_{mn} &\equiv& [\cD_m, \cD_n],~~~~\cFb_{mn} \equiv [\cDb_m, \cDb_n],
\eea
with the complexified covariant derivatives: $\cD_m \equiv \partial_m + [\cA_m, \cdot~],~~\cDb_m \equiv \partial_m + [\cAb_m, \cdot~]$.

The auxiliary field $d$ in the above action can be integrated out using the equation of motion
\beq
d = -D_m \phi_m = \frac{i}{2}[\cD_m, \cDb_m].
\eeq

The supersymmetry charges of the original theory also undergo a decomposition similar to that of the fermions, resulting in scalar, vector and anti-symmetric tensor supercharges, $\cQ$, $\cQ_m$ and $\cQ_{mn}$, respectively. The twisted supersymmetry algebra contains a subalgebra in which the scalar supercharge is strictly nilpotent: $\cQ^2 = 0$. Since this subalgebra does not generate any translations we can easily transport the theory on to the lattice by preserving one supersymmetry charge exact at finite lattice spacing.

The twisted action can be written as a sum of $\cQ$-exact and $\cQ$-closed terms
\beq
\label{eq:5d_n2_action}
S_5 = \cQ \Lambda - \frac{1}{g_5^2} \int d^5x ~\Tr \frac{i}{8} \epsilon_{mncde} \chi_{de} \cDb_c \chi_{mn},
\eeq
where $\Lambda$ is a functional of the fields
\beq
\Lambda = \frac{1}{g_5^2}  \int d^5x ~\Tr \Big(\frac{i}{4} \chi_{mn} \cF_{mn} - \eta D_m \phi_n - \hf \eta d \Big).
\eeq

It is easy to see that the twisted action is $\cQ$-invariant. The $\cQ$-exact piece vanishes trivially due to the nilpotent property of the scalar supercharge and the vanishing of $\cQ$-closed piece can be shown through the Bianchi identity for covariant derivative
\beq
\epsilon_{mncde} \cDb_c\cFb_{de} = \epsilon_{mncde} [\cDb_c,[\cDb_d, \cDb_e]] = 0.
\eeq  

The supersymmetry transformations generated by the scalar supercharge have the following form
\begin{align}
 \cQ \cA_m  		&= \psi_m,   			&	\cQ \psi_m	&= 0,	&	\cQ \cAb_m	&= 0, \\
 \cQ \chi_{mn} 	&= -i \cFb_{mn}, 	&	\cQ \eta		&= d,	&	\cQ d				&= 0,
\end{align}
and it is easily seen that the $\cQ$ supersymmetry charge is strictly nilpotent on the twisted fields.
 
\section{Lattice Construction of 5D MSYM}
\label{subsec:sym-lattice-n2d5}

It is straightforward to discretize the five-dimensional MSYM theory once the action is written in the twisted form. One has to address several technicalities once the theory is formulated on the lattice. The lattice construction is provided in detail in Ref. \cite{Joseph:2016tlc}. We consider a hypercubic lattice made out of unit cells containing five mutually orthogonal basis vectors $\hatbmu_m$ in the positive $x_m$ directions
\begin{align}
 \hatbmu_1	&= (1, 0, 0, 0, 0),	& \hatbmu_2	&= (0, 1, 0, 0, 0),	&	\hatbmu_3	&= (0, 0, 1, 0, 0), \nn \\
 \hatbmu_4	&= (0, 0, 0, 1, 0),	& \hatbmu_5	&= (0, 0, 0, 0, 1).	&	\nn
\end{align}

In the bosonic sector, this theory has only a complexified vector field with five components. The complex continuum gauge fields $\cA_m$, $m=1, \cdots, 5$, are mapped to complexified Wilson gauge links $\cU_m(\bn)$ and they are placed on the links connecting lattice site $\bn$ to site $\bn + \hatbmu_m$ of the hypercubic lattice. The field $\cUb_m(\bn)$ is placed on the oppositely oriented link, that is, from site $\bn + \hatbmu_m$ to site $\bn$. The complexified field strength is defined in the following way on the lattice
\beq
\cF_{mn}(\bn) \equiv \cD_m^{(+)}\cU_n(\bn) = \cU_m(\bn)\cU_n(\bn + \hatbmu_m) - \cU_n(\bn)\cU_m(\bn + \hatbmu_n).
\eeq
We map the complexified covariant derivative $\cD_m$ into a forward or backward lattice covariant difference operator, $\cD^{(+)}_m$ or $\cD^{(-)}_m$. The discretization rules appropriate for twisted supersymmetric gauge theories are derived in Ref. \cite{Damgaard:2007be}.

The Grassmann-odd fields of the twisted theory have the interpretation as geometric fermions \cite{Rabin:1981qj}. Since the fermions of the twisted theory are p-forms (p $= 0,1,2$), it is natural to place each of them on the p-cell of the five-dimensional hypercubic lattice.

The nilpotent scalar supersymmetry charge acts on the lattice fields in the following way
\begin{align}
 \cQ \cU_m(\bn)	&= \psi_m(\bn),	&	\cQ \cUb_m(\bn)	&= 0, & \cQ \psi_m(\bn)			&= 0, \nn \\
 \cQ \eta(\bn)	&= d(\bn),			& \cQ d(\bn)			&= 0,	&	\cQ \chi_{mn}(\bn)	&= -i \Big(\cDb_m^{(+)}\cUb_n(\bn)\Big) = -i \cFb_{mn}^L.  \nn
\end{align}

We need to ensure that the placements of the fields on the hypercubic lattice results in a gauge invariant lattice action. The mappings and orientations of the lattice variables can be easily summarized by providing their gauge transformation properties on the lattice. For $g(\bn)$, a unitary matrix at lattice site $\bn$, which is an element of the gauge group $G$, we have the gauge transformations on the lattice fields
\begin{align}
 \label{eq:gauge-t}
 \cU_m(\bn) 			&\rightarrow g(\bn) \cU_m(\bn) g^{\dagger}(\bn + \hatbmu_m),  &	\cUb_m (\bn) 	&\rightarrow g(\bn + \hatbmu_m) \cUb_m(\bn) g^{\dagger}(\bn), \nn \\
 \eta(\bn) 				&\rightarrow g(\bn) \eta(\bn) g^{\dagger}(\bn),   						&	\psi_m(\bn) 	&\rightarrow g(\bn) \psi_m(\bn) g^{\dagger}(\bn + \hatbmu_m), \\
 \chi_{mn} (\bn) 	&\rightarrow g(\bn + \hatbmu_m + \hatbmu_n) \chi_{mn}(\bn) g^{\dagger}(\bn).  &    \nn
\end{align}

It is clear from above that the gauge transformations depend on the geometric nature of the lattice fields we are considering.   

The covariant forward and backward difference operators act on the lattice fields the following way \cite{Damgaard:2007be}
\bea
\label{eq:discretization-1}
\cD_m^{(+)} f(\bn) &=& \cU_m(\bn) f(\bn + \hatbmu_m) - f(\bn) \cU_m(\bn), \\
\cD_m^{(+)} f_n(\bn) &=& \cU_m(\bn) f_n(\bn + \hatbmu_m) - f_n(\bn) \cU_m(\bn + \hatbmu_n), \\
\cDb_m^{(-)} f_m(\bn) &=& f_m(\bn)\cUb_m(\bn) - \cUb_m(\bn - \hatbmu_m) f_m(\bn - \hatbmu_m), \\
\label{eq:discretization-2}
\cDb_c^{(+)} f_{mn} (\bn) &=& f_{mn}(\bn + \hatbmu_c) \cUb_c(\bn) - \cUb_c(\bn + \hatbmu_m + \hatbmu_n) f_{mn}(\bn).
\eea

It is now straightforward to write down the lattice action of the five-dimensional MSYM. After integrating out the auxiliary field $d$ using its equation of motion
\beq
d(\bn) = -\frac{i}{2} \sum_m \cDb^{(-)}_m \cU_m(\bn),
\eeq
the $\cQ$-exact piece of the action takes the following form
\bea
S_{\cQ-\textrm{exact}} &=& \beta \sum_{\bn, m, n} \Tr \Big(- \qtr \cFb_{mn}^L(\bn) \cF_{mn}(\bn) - \frac{1}{8} \Big(\cDb_m^{(-)}\cU_m(\bn)\Big)^2 \nn \\
&&\quad \quad \quad \quad \quad \quad \quad \quad - i\chi_{mn}(\bn) \cD^{(+)}_m\psi_n(\bn) - i\eta(\bn) \cDb^{(-)}_m\psi_m(\bn) \Big),
\eea
with $\beta$ denoting the lattice coupling. In terms of the `t Hooft coupling $\lambda$ and lattice spacing $a$ it is
\beq
\beta = \frac{N}{\lambda},~~\lambda = \frac{N g_5^2}{a}.
\eeq

Following the set of prescriptions given in Eq. (\ref{eq:gauge-t}) we see that the $\cQ$-exact piece of the action is gauge invariant; each term forms a closed loop on the lattice. The $\cQ$-closed term needs special consideration. It must be modified on the lattice in order to maintain gauge invariance. In order to make it gauge invariant we introduce the ordered product of link variables along a path connecting lattice sites $\bn$ and $\bn + \hatbmu_m + \hatbmu_n + \hatbmu_c + \hatbmu_d + \hatbmu_e$ \cite{Joseph:2016tlc}. Let $C_L$ be a path on the lattice connecting these two sites. Then the path ordered link (POL) is defined as
\beq
\cP_{\rm POL} \equiv \prod_{l \in C_L} \cUb_l, 
\eeq
with $\cUb_l$ denoting a link variable on $C_L$. We choose $\cUb$-fields to form the path ordered link since it is trivially annihilated by $\cQ$-supersymmetry. If one chooses to form the path ordered link using $\cU$-fields the term becomes gauge invariant but it breaks the most important property of the lattice theory, that is, $\cQ$-symmetry at the lattice level.

The $\cQ$-closed term on the lattice now takes the following gauge invariant form
\bea
S_{\cQ-\rm closed} &=& - \frac{i\beta}{8} \sum_{\bn, m,n,c,d,e} \Tr \epsilon_{mncde} \cP_{\rm POL} \chi_{de}(\bn + \hatbmu_m + \hatbmu_n + \hatbmu_c) \cDb^{(+)}_c \chi_{mn}(\bn).~~~~~~~~
\eea

We also note that the above $\cQ$-closed term reduces to its continuum counterpart in the naive continuum limit in which the gauge links are set to unity. 

The lattice action constructed here is invariant under $\cQ$-supersymmetry. The $\cQ$-variation of the $\cQ$-exact term vanishes due to the property $\cQ^2 = 0$. The $\cQ$-closed term vanishes due to Bianchi identity for covariant derivatives on the lattice
\beq
\epsilon_{decmn} \cDb^{(+)}_c \cFb_{mn}^L = 0,
\eeq
with $d, e = 1, \cdots, 5$. It can also be shown that the lattice theory constructed here is free from spectrum doublers \cite{Joseph:2016tlc} using the tools developed in Ref. \cite{Catterall:2011pd}.

\section{Discussion and Conclusions}
\label{sec:conclusion}

We have constructed a supersymmetric lattice action of the five-dimensional MSYM theory using the methods of topological twisting and geometric discretization. The lattice theory preserves one supersymmetry charge exact at finite lattice spacing. The covariant derivatives of the continuum theory are mapped to forward and backward covariant difference operators through a well defined prescription. The lattice theory is supersymmetric, gauge invariant and free from spectrum doublers. 

The lattice formulation proposed here can be used to explore the non-perturbative regime of MSYM theories in five dimensions at finite gauge coupling and number of colors, and also its parent theory, $(2, 0)$ superconformal theory in six dimensions. It would be interesting to find a nontrivial UV fixed point from the lattice theory for 5D MSYM since the fixed point can give a UV completion and non-perturbative definition of the theory.   

The theory described here takes part in the AdS-CFT correspondence. It is the low energy theory living on a stack of D4 branes. The gravitational dual of this theory is known; it is the supergravity on the near horizon geometry of D4 branes.

\section{Acknowledgments}
\label{sec:ack}

I thank Joel Giedt, Masanori Hanada, David Schaich and David B. Kaplan for fruitful discussions on lattice supersymmetry during LATTICE 2016. I also thank Apoorva Patel for enlightening discussions on the lattice formulation presented here during the visit to Indian Institute of Science. This work was supported in part by the European Research Council under the European Union's Seventh Framework Programme (FP7/2007-2013), ERC grant agreement STG 279943, ``Strongly Coupled Systems".


\begin{thebibliography}{99}

\bibitem{Douglas:2010iu} 
  M.~R.~Douglas,
  ``On D=5 super Yang-Mills theory and (2,0) theory,''
  JHEP {\bf 1102}, 011 (2011)
  [arXiv:1012.2880 [hep-th]]; N.~Lambert, C.~Papageorgakis and M.~Schmidt-Sommerfeld,
  ``M5-Branes, D4-Branes and Quantum 5D super-Yang-Mills,''
  JHEP {\bf 1101}, 083 (2011)
  [arXiv:1012.2882 [hep-th]].

\bibitem{Tachikawa:2011ch} 
  Y.~Tachikawa,
  ``On S-duality of 5d super Yang-Mills on $S^1$,''
  JHEP {\bf 1111}, 123 (2011)
  doi:10.1007/JHEP11(2011)123
  [arXiv:1110.0531 [hep-th]].

\bibitem{Kallen:2012cs} 
  J.~Kallen and M.~Zabzine,
  ``Twisted supersymmetric 5D Yang-Mills theory and contact geometry,''
  JHEP {\bf 1205}, 125 (2012)
  doi:10.1007/JHEP05(2012)125
  [arXiv:1202.1956 [hep-th]].

\bibitem{Kallen:2012va} 
  J.~Kallen, J.~Qiu and M.~Zabzine,
  ``The perturbative partition function of supersymmetric 5D Yang-Mills theory with matter on the five-sphere,''
  JHEP {\bf 1208}, 157 (2012)
  doi:10.1007/JHEP08(2012)157
  [arXiv:1206.6008 [hep-th]].

\bibitem{Kim:2012ava} 
  H.~C.~Kim and S.~Kim,
  ``M5-branes from gauge theories on the 5-sphere,''
  JHEP {\bf 1305}, 144 (2013)
  doi:10.1007/JHEP05(2013)144
  [arXiv:1206.6339 [hep-th]].

\bibitem{Kallen:2012zn} 
  J.~Kallen, J.~A.~Minahan, A.~Nedelin and M.~Zabzine,
  ``$N^3$-behavior from 5D Yang-Mills theory,''
  JHEP {\bf 1210}, 184 (2012)
  [arXiv:1207.3763 [hep-th]]. 

\bibitem{Bern:2012di}
  Z.~Bern, J.~J.~Carrasco, L.~J.~Dixon, M.~R.~Douglas, M.~von Hippel and H.~Johansson,
  ``D=5 maximally supersymmetric Yang-Mills theory diverges at six loops,''
  Phys.\ Rev.\ D {\bf 87} (2013) 2,  025018
  [arXiv:1210.7709 [hep-th]]. 

\bibitem{Minahan:2013jwa} 
  J.~A.~Minahan, A.~Nedelin and M.~Zabzine,
  ``5D super Yang-Mills theory and the correspondence to AdS$_7$/CFT$_6$,''
  J.\ Phys.\ A {\bf 46}, 355401 (2013)
  doi:10.1088/1751-8113/46/35/355401
  [arXiv:1304.1016 [hep-th]].

\bibitem{Klebanov:1996un} 
  I.~R.~Klebanov and A.~A.~Tseytlin,
  ``Entropy of near extremal black p-branes,''
  Nucl.\ Phys.\ B {\bf 475}, 164 (1996)
  [hep-th/9604089]; M.~Henningson and K.~Skenderis,
  ``The Holographic Weyl anomaly,''
  JHEP {\bf 9807}, 023 (1998)
  [hep-th/9806087]. 

\bibitem{Maldacena:1997re} 
  J.~M.~Maldacena,
  ``The Large N limit of superconformal field theories and supergravity,''
  Int.\ J.\ Theor.\ Phys.\  {\bf 38}, 1113 (1999)
  [Adv.\ Theor.\ Math.\ Phys.\  {\bf 2}, 231 (1998)]
  [hep-th/9711200]. 

\bibitem{Lattice_Susy}   
  D.~Kadoh, ``Precision test of the gauge/gravity duality in two-dimensional $\cN=(8,8)$ SYM," PoS(LATTICE2016)033; 
	V.~Forini, ``Strings on the lattice and AdS/CFT," PoS(LATTICE2016)206; 
	J.~Giedt, ``S-duality in lattice N=4 super Yang-Mills," PoS(LATTICE2016)209;
	S.~Kamata, ``Numerical Analysis of Discretized $\cN =(2,2)$ SYM on Polyhedra," PoS(LATTICE2016)210; 
	D.~Schaich, ``Latest results from lattice $\cN = 4$ supersymmetric Yang--Mills," PoS(LATTICE2016)221; 
	P.~Giudice, ``Simulations of $\cN=1$ supersymmetric Yang-Mills theory with three colours," PoS(LATTICE2016)222;
	D.~August, ``Spectroscopy of two dimensional $\cN=2$ Super Yang Mills theory," PoS(LATTICE2016)234;
	E.~Berkowitz, ``Supergravity from Gauge Theory," PoS(LATTICE2016)238, arXiv:1608.01951 [hep-lat].
  
\bibitem{Geyer:2002gy} 
  B.~Geyer and D.~Mulsch,
  ``Higher dimensional analog of the Blau-Thompson model and N(T) = 8, D = 2 Hodge type cohomological gauge theories,''
  Nucl.\ Phys.\ B {\bf 662}, 531 (2003)
  [hep-th/0211061].    
  
\bibitem{Joseph:2016tlc} 
  A.~Joseph,
  ``A Euclidean Lattice Formulation of D=5 Maximally Supersymmetric Yang-Mills Theory,''
  JHEP {\bf 1606}, 030 (2016)
  doi:10.1007/JHEP06(2016)030
  [arXiv:1604.02707 [hep-lat]].  
  
\bibitem{Damgaard:2007be} 
  P.~H.~Damgaard and S.~Matsuura,
  ``Classification of supersymmetric lattice gauge theories by orbifolding,''
  JHEP {\bf 0707}, 051 (2007)
  [arXiv:0704.2696 [hep-lat]]; S.~Catterall,
  ``From Twisted Supersymmetry to Orbifold Lattices,''
  JHEP {\bf 0801}, 048 (2008)
  doi:10.1088/1126-6708/2008/01/048
  [arXiv:0712.2532 [hep-th]]; P.~H.~Damgaard and S.~Matsuura,
  ``Geometry of Orbifolded Supersymmetric Lattice Gauge Theories,''
  Phys.\ Lett.\ B {\bf 661}, 52 (2008)
  [arXiv:0801.2936 [hep-th]].
  
\bibitem{Rabin:1981qj} 
  J.~M.~Rabin,
  ``Homology Theory of Lattice Fermion Doubling,''
  Nucl.\ Phys.\ B {\bf 201}, 315 (1982); P.~Becher and H.~Joos,
  ``The Dirac-Kahler Equation and Fermions on the Lattice,''
  Z.\ Phys.\ C {\bf 15}, 343 (1982); T.~Banks, Y.~Dothan and D.~Horn,
  ``Geometric Fermions,''
  Phys.\ Lett.\ B {\bf 117}, 413 (1982). 
  
\bibitem{Catterall:2011pd} 
  S.~Catterall, E.~Dzienkowski, J.~Giedt, A.~Joseph and R.~Wells,
  ``Perturbative renormalization of lattice N=4 super Yang-Mills theory,''
  JHEP {\bf 1104}, 074 (2011)
  doi:10.1007/JHEP04(2011)074
  [arXiv:1102.1725 [hep-th]].   

\end{thebibliography}
\end{document}